\RequirePackage[2020-02-02]{latexrelease}
\documentclass[10pt]{revtex4}
\usepackage{graphicx}
\usepackage{amsmath,amssymb}
\usepackage{mathtools}
\newcommand{\ket}[1]{\left | #1 \right\rangle}

\begin{document}

\title{Interference in complex canonical variables is not quantum}
\author{Chiara Marletto and Vlatko Vedral}
\affiliation{Clarendon Laboratory, University of Oxford, Parks Road, Oxford OX1 3PU, United Kingdom}

\begin{abstract}
We formally represent the quantum interference of a single qubit embodied by a photon in the Mach-Zehnder interferometer using the classical Hamiltonian framework but with complex canonical variables. Although all operations on a single qubit can be formally expressed using the complex classical Hamiltonian dynamics, we show that the resulting system is still not a proper qubit. The reason is that it is not capable of getting entangled to another bona fide qubit and hence it does not have the information-processing capacity of a fully-fledged quantum system. This simple example powerfully illustrates the failure of all hybrid quantum-classical models in accounting for the full range of behaviour of even a single quantum bit. 
\end{abstract}

\maketitle

Hamilton's formulation of Newton's laws relies on the use of the canonically conjugate position $q$ and momentum $p$. The equations of motion follow from the Hamiltonian $H=H(q,p)$ and are 
\begin{equation} 
\frac{\partial q}{\partial t} = \frac{\partial H}{\partial p} \;\;\;\;\; \frac{\partial p}{\partial t} =- \frac{\partial H}{\partial q}
\end{equation}
In the case of a free particle of unit mass, $H=p^2/2$ and the above equations give us $\dot q = p$ and $\dot p = 0$. These are easily solved and lead to the usual equation for inertial motion $q(t) = q(0)+pt$. 

In this paper we will be dealing with harmonic oscillators, the reason for which will become apparent below. At present it suffices to say that the Hamiltonian is given by $H=p^2/2+q^2/2$ where, without any loss of generality, we have assumed that both the mass and frequency are unit. The equations of motion are now $\dot q = p$ and $\dot p = -q$, which are just two coupled first order differential equations. This reduces to $\ddot q + q = 0$ and $\ddot p + p = 0$. 

The key trick we use in this paper is to define a new set of complex coordinates: $z=(q+ip)/\sqrt{2}$ and $z^*=(q-ip)/\sqrt{2}$ \cite{Strocchi}. Hamilton's equations of motion now read
\begin{equation} 
i\frac{\partial z}{\partial t} = \frac{\partial H}{\partial z^*} \;\;\;\;\; -i\frac{\partial z^*}{\partial t} =- \frac{\partial H}{\partial z}
\end{equation}
In other words, $z$ and $z^*$ are treated as independent variables, just as $q$ and $p$. The Hamiltonian for the harmonic oscillator is now even simpler $H(z,z^*) = zz^*$ and the equations of motion for the complex coordinates are: $i\dot z = z$ and $-i\dot z^* = z^*$. 

In general, it is clear that the complex representation is completely equivalent to the original one based on the canonical position and momentum. We would now like to show that the complex classical Hamiltonian description can account for the phenomenon of single qubit interference. 

A single photon going through a Mach-Zehnder interferometer has been a foremost way of thinking about interference in quantum information and computation \cite{Nielsen}. It is analogous to the double-slit experiment, which in the words of Feynman contains ``the only mystery" in quantum physics \cite{Feynman}. Consider $a_x$ and $a_x^{\dagger}$, the bosonic annihilation and creation operators for photons in mode $x$ (see e.g. \cite{Heitler} for a general treatment of the quantised electromagnetic field), with the property that $[a_x, a_y^{\dagger}]=\delta_{x,y}$, $[a_x, a_y]=0$, $a_x\ket{0}=0\;\forall x$, where $\ket{0}$ is the vacuum state of the global Fock space and $[A,B]=AB-BA$ is the commutator of the two operators $A$ and $B$. We will assume that $x$ represents a region of space where the photon can be confined to arbitrarily high accuracy; in this case, $x$ can be either $L$ (a region around the left arm of the interferometer) or $R$ (a region around the right arm of the interferometer). These two regions are non-overlapping (their separation being much larger than their respective extents). All these assumptions hold in practice. 

In the Schr\"odinger picture, the quantum state of the photon (we shall not go into the subtleties about how to define it as they are not relevant for the present discussion) changes after the first beamsplitter ($U_{BS}$), then acquires the same phase shift in both arms (which can therefore be ignored), and finally undergoes another change at the final beamsplitter. Labelling the quantum state where the photon is on the left or right arm of the interferometer respectively as $\ket{L}\doteq a_{L}^{\dagger}\ket{0}$ and $\ket{R}\doteq a_{R}^{\dagger}\ket{0}$, the dynamical evolution of the photon is given by:
\begin{eqnarray}
|L\rangle \xrightarrow[\text{}]{U_{BS}} \frac{1}{\sqrt{2}}(\ket{L}+\ket{R}) \xrightarrow[\text{}]{U_{BS}} |L\rangle \; ,
\end{eqnarray}
In terms of operators, we have started in an eigenstate of $Z=a_{L}^{\dagger}a_L-a_{R}^{\dagger}a_R$, then changed to $X=a_{L}^{\dagger}a_R+a_{R}^{\dagger}a_L$ and then finally rotated back into $Z$. 

We will now show that the same formal account can be given with the complex classical description. We need the conjugate complex numbers to represent the two modes $z_L,z^*_L$ and $z_R,z^*_R$. The Hamiltonian is given by $H= z_L z^*_L + z_R z^*_R$. Our choice of the initial vales for the complex coordinates should be $z_L=1=z^*_L$ and $z_R=0=z^*_R$ in order to model a single photon in the $L$ mode. The Hadamard gate is then represented by the transformation
\begin{equation}
z_L\rightarrow \frac{z_L+z_R}{\sqrt{2}} \;\;\; z_R\rightarrow \frac{z_L-z_R}{\sqrt{2}}
\end{equation} 
and the corresponding transformations for the complex conjugates. This transforms the initial $Z= z_L z^*_L - z_R,z^*_R$ into $X=z_L z^*_R + z_R,z^*_L$. It is clear that another application of the Hadamard gate takes us back to $Z$ which is the same as the quantum sequence. In fact, we could model this by introducing a classical state written as a two-vector $(z_L,z_R)$.  The initial state is then $(1,0)$, which, after the action of the first Hadamard transformation becomes $(1/\sqrt{2},1/\sqrt{2})$, mimicking the quantum superposition across the two modes. The second Hadamard then takes us back to the original state exactly as in the quantum treatment. 

Note also that any phase gate introduces in either of the modes can now be accounted for classically because we allow complex numbers. It would simply be represented by the transformation $z \rightarrow e^{i\theta}z$. This is the same effect that the annihilation operator in the quantum treatment would suffer. In this way, the component $Y= i(z_L z^*_R - z_R,z^*_L)$ can also be reached. It is, therefore, shown that the full set of transformations on a single qubit can be represented with complex classical Hamiltonian mechanics. 

{However, despite this formal analogy, the classical complex bit is still not the same as the qubit. In fact, the process described above is not an actual interference experiment. What counts as an interference experiment is determined by what dynamical variables are actually observables. The complex bit and the qubit differ radically in this regard. This is because, crucially}, the classical complex bit does not obey the uncertainty principle between different $x,y,z$ components. In the case of a qubit these obey the algebra of Pauli matrices and one component being sharp exactly leaves us with the maximum uncertainty in the other two. This is not true of the classical complex bit for which the values in all three directions can be sharp at the same time (as in the case of the interferometer above).  

A different way of expressing the above fact is that the classical complex bit cannot be entangled with another qubit. The notion of entanglement simply does not exist in the hybrid system consisting of one complex and one quantum bit. Intuitively speaking, if it did exist, we could faithfully transfer the information from the qubit into the complex bit thereby being able to violate the uncertainty principle on the qubit by measuring the complex bit (whose components can all be specified at the same time). The complex bit does not and could not ``understand" how to respond to a superposition state of the qubit. This is because the complex bit is an information media (in the language of constructor theory, \cite{DEUMA}), whose observables can all be copied simultaneously to arbitrarily high accuracy by the same measuring device; while a qubit is a superinformation medium, whose different components cannot be copied in the same manner. 

As a way of illustrating this, suppose that we want to swap the states of the complex and quantum bits. The Hamiltonian for this operation would then be written as $H= (z_L z^*_R + z_R,z^*_L)X + (i(z_L z^*_R - z_R,z^*_L))Y$, where now $X$ and $Y$ are the Pauli qubit operators. This Hamiltonian, however, would only modify the state of the qubit in proportion to the (real) coefficients specified by the corresponding representations of $X$ and $Y$ on the complex bit. The states would thus not be swapped. All semi-classical models suffer from this and related problems. 

The fact that a complex bit can describe the single photon interference in an interferometer should not come as a surprise. We know that Maxwell's equations can be cast in the form of a quantum-like equation with the quantity $\bf{E}+i\bf{B}$ playing the role of the wavefunction \cite{Birula}. Single photon interference is therefore perfectly well captured by the classical theory even though the concept of a photon does not exist there. However, there is no sense in which this description of a photon can ever be used to model the entanglement of the kind that, for instance, occurs in the Jaynes-Cummings model between a qubit and a single mode of light \cite{Jaynes}.  

We hope that this short exposition sheds some light on the hotly debated topic of hybrid systems \cite{Terno} in which one subsystem is bone fide quantum and the other classical. While we know that such hybrid systems fail to capture most of quantum electrodynamics, it is still an open question what happens in quantum gravity. Here we need further experimental evidence to guide us in the right direction for, despite the fact that the hybrid models are theoretically inconsistent, nature may well have chosen them in some domain of its inner workings.

\textit{Acknowledgments}: This research was made possible by the generous support of the Gordon and Betty Moore Foundation, the Eutopia Foundation, and the John Templeton Foundation, as part of The Quantum Information Structure of Spacetime (QISS) Project (qiss.fr). The opinions expressed in this publication are those of the authors and do not necessarily reflect the views of the John Templeton Foundation.

\end{document}